\documentclass{article}
\usepackage{amsmath,amsfonts,amsthm,amssymb,amscd,color,xcolor,mathrsfs}
\usepackage{amsfonts}
\usepackage{amsmath,amsthm}
\usepackage{hyperref}
\usepackage{latexsym}
\usepackage{array}
\usepackage{amssymb}
\usepackage{enumerate}

\usepackage[francais,english]{babel}

\usepackage{color}
\usepackage[latin1]{inputenc}

\DeclareFontFamily{U}{wncy}{}
    \DeclareFontShape{U}{wncy}{m}{n}{<->wncyr10}{}
    \DeclareSymbolFont{mcy}{U}{wncy}{m}{n}
    \DeclareMathSymbol{\Sh}{\mathord}{mcy}{"58}

\theoremstyle{plain}
\newtheorem{theorem}{Theorem}[section]
\newtheorem*{theorem*}{Theorem}
\newtheorem{lemma}[theorem]{Lemma}
\newtheorem{proposition}[theorem]{Proposition}
\newtheorem{corollary}[theorem]{Corollary}

\theoremstyle{remark}

\newtheorem*{lem*}{Lemma}
\newtheorem*{sublem*}{Sublemma}
\newtheorem*{remark*}{Remark}
\newtheorem*{NB*}{NB}

\newcommand{\Sv}{ \Sigma^{nbh}}
\newcommand{\Svm}{ \Sigma^{nbh_m}}
\newcommand{\Sod}{ \Sigma^1}
\newcommand{\nbh}{\text{neighbourhood}}

\newcommand{\thb}{ \bar\theta }

\newcommand{\R}{ \mathbb{R} }

\newcommand{\cI}{ \mathcal{I}}

\newcommand{\cC}{ \mathcal{C} }

\newcommand{\cN}{ \mathcal{N} }

\newcommand{\cJ}{ \mathcal{J} }

\newcommand{\om}{ \omega }

\newcommand{\G}{ \Gamma }
\renewcommand{\phi}{ \varphi }

\newcommand{\eps}{\varepsilon}

\newcommand{\la}{ \lambda }

\newcommand{\meas}{\operatorname{meas}}

\newcommand{\dist}{{\operatorname{dist}}}

\newcommand{\be}{\begin{equation}}
\newcommand{\ee}{\end{equation}}
\newcommand{\ben}{\begin{equation*}}
\newcommand{\een}{\end{equation*}}

\numberwithin{equation}{section}

\newcommand{\lan}{ \langle }

\newcommand{\ran}{ \rangle}

\newcommand{\p}{ \partial}

\author{Sergei Kuksin\footnote{CNRS, Institut de Math\'emathiques de Jussieu--Paris Rive Gauche, UMR 7586, Universit\'e Paris Diderot, Sorbonne Paris Cit\'e, F-75013, Paris, France; e-mail: \href{mailto:Sergei.Kuksin@imj-prg.fr}{Sergei.Kuksin@imj-prg.fr}} 
}

\title
{Asymptotic  expansions for some  integrals of quotients with degenerated divisors }

\begin{document}

\maketitle

\begin{abstract}

We study asymptotic expansion as $\nu\to0$ for integrals over ${ \mathbb{R} }^{2d}=\{(x,y)\}$ of quotients 
$F(x,y) \big/ \big( (x\cdot y)^2+(\nu \Gamma(x,y))^2\big)$, where $\Gamma$ is strictly positive and $F$ decays 
at infinity sufficiently fast. Integrals of this kind appear in description of the four--waves interactions. 

 \end{abstract}


\section{Introduction}\label{s2}

   Our concern is the integrals 
\be\label{I_s}
\begin{split}
I_\nu=
 \int_{\R^d\times\R^d}  dx\,dy\, \,   \frac{ F(x,y)}{ (x\cdot y)^2 +(\nu \Gamma(x, y))^2}\,,\quad d\ge2\,,\;
 0<\nu\ll	1.
\end{split}
\ee
 Such  integrals and their singular limits $\nu\to0$ 
  appear in   physical works on the four-waves 
 interaction, where the latter is suggested  as a mechanism, dictating the 
 long-time behaviour of solutions for nonlinear Hamiltonian PDEs
 with cubic nonlinearities and  large values of the space-period. Usually  the integrals $I_\nu$
 appear there in an implicit form, and become visible as a result of rigorous 
 mathematical analysis of the objects and constructions, involved in the heuristic physical  argument
 (see below in this section).

We denote $(x,y)=z\in\R^{2d}$, $\omega(z)=x\cdot y$, and assume that 
$F$ and $\Gamma$ are $C^2$--smooth real 
functions, satisfying \footnote{For example, $\Gamma = \lan z\ran^{2m}$, $m\in\R$,  and $F$ is a Schwartz function.} 
\be\label{F_1}
 |\p_z^\alpha F(z)|\le K \lan z \ran^{-M-|\alpha|}\,\quad \forall\,z,\;
 \forall\, |\alpha|\le 2\,;
\ee
\be\label{Ga_1}
\begin{split}
 |\Gamma(z)|\ge K^{-1}  \lan z \ran^{r_*}\;\;\forall\,z\,,  \qquad 
|\p_z^\alpha \Gamma(z)|\le K  \lan z \ran^{r_* -|\alpha|}\;\;  \forall\,z,\;
\forall\, |\alpha|\le 2\,.
\end{split}
\ee
Here
 $r_*, M, K$ are any real  constants such that
\be\label{hr}
 M+r_*>2d- 2, \; \quad M>2d-4,\
 \quad K>1\,.
\ee
As usual we denote $\lan z\ran = \sqrt{|z|^2+1}$.

 The main difficulty in the study of $I_\nu$ comes from the vicinity of  the quadric  $\Sigma =\{\om(z)=0\}$. 
 The latter  has a locus at $0\in\R^{2d}$ and is smooth outside it. Firstly  we will study $I_\nu$ near $0$, next -- near 
 the smooth part of  the quadric, $\Sigma\setminus\{0\} = : \Sigma_*$, and finally will combine the results obtained 
 to get the main result of this work:

  \begin{theorem}\label{t_singint}
  As $\nu\to0$, the  integral $ I_\nu$ has the  following asymptotic: 
  \be\label{p22}
   I_\nu  = \pi \nu^{-1} \int_{\Sigma_*} \frac{F(z)}{|z|\, \Gamma(z)}\, d_{\Sigma_*} z
   + I^\Delta. 
  \ee
  Here 
   $d_{\Sigma_*} z$ is the volume element on $\Sigma_*$ 
   and    $
   |I^\Delta|\le C \,  \chi_d(\nu),
$
where 
 \be\label{chi_d}
 \chi_d(\nu)=
   \left\{\begin{array}{ll}
 1,& d\ge3 \,,
 \\
 \max(1,\ln(\nu^{-1}) ),& d=2\,.
\end{array}\right.
\ee
 The integral in \eqref{p22} converges absolutely, and the constant  $C$ depends on 
  $d, K, M$ and $r_*$.
 \end{theorem}

  The integral in \eqref{p22} may be regarded as  integrating  of  the function $F/\G$ 
 against  a 
 measure in  $\R^{2d}$, supported by $\Sigma$, which  we will  denote  $|z|^{-1} \delta_{\Sigma_*}$
 (here $\delta_{\Sigma_*}$ is the delta--function of the hypersurface $\Sigma_*$). 
   For any real number $m$ let  $C_m(\R^{2d})$ be the space of continuous functions on 
   $\R^{2d}$ with finite norm  $|f|_m = \sup_z|f(z)| \lan z\ran^m$. By \eqref{F_1} and \eqref{Ga_1}, 
   $F/\G\in C_{M+r_*}(\R^{2d})$, where  $M+r_*>2d-2$. 
 
 \begin{proposition}\label{prop}
 The measure  $|z|^{-1} \delta_{\Sigma_*}$ is an atomlesss $\sigma$-finite Borel measure on $\R^{2d}$.
 The integrating over it 
 defines a continuous linear functional on the space $C_m(\R^{2d})$ if $m>2d-2$.
 \end{proposition}
 
 Since any function $F\in C^\infty_0(\R^{2d})$ satisfies \eqref{F_1} for every $M$, 
 then we have 
 
 \begin{corollary}
 Let a $C^2$--function $\G$ meets \eqref{Ga_1} with  some $r_*$.  Then
  the function 
 $
{ \nu}/ {\big( (x\cdot y)^2 +(\nu \Gamma(x, y))^2\big)}
$
converges to the measure $ |z|^{-1}\delta_{\Sigma_*}$ as $\nu\to0$, in the space of distributions.
 \end{corollary}

  The theorem and the proposition are proved below in Sections \ref{s_71}--\ref{s4}.
  \medskip

 In the mentioned above works from the non-linear physics,   to describe  the long-time behaviour of solutions for nonlinear  Hamiltonian PDEs with cubic nonlinearities,
 physicists derived  nonlinear kinetic equations,  called the (four-) 
 wave kinetic equations. The $k$-th component of the kinetic kernel $K$ ($k\in\R^d$) for such  equation 
 is given by an integral of the following form:
 \be\label{heur}
 K_k=
 \int_{\R^d} \int_{\R^d} \int_{\R^d} F_k(k_1, k_2, k_3) \delta^{k k_3}_{k_1 k_2}
 \delta(\omega^{k k_3}_{k_1 k_2})\, dk_1 dk_2 dk_3\,.
 \ee
 Here $\delta^{k k_3}_{k_1 k_2}$ is the delta-function $\{k+k_3 = k_1+k_2\}$ and 
 $\delta(\omega^{k k_3}_{k_1 k_2})$ is the delta-function $\{\omega_k+\omega_{k_3} = 
 \omega_{k_1}+\omega_{k_2}\}$, where $\{\omega_k\}$ is the spectrum of  oscillations 
 for the linearised at zero equation. If the corresponding nonlinear PDE is the cubic NLS
 equation, then $\omega_k = |k|^2$. In this case the two delta-functions define the following
 algebraic  set:
 $$
 \{(k_1, k_2, k_3)\in \R^{3d}: k+k_3= k_1+k_2,\;\; |k|^2+ |k_3|^2= |k_1|^2+ |k_2|^2\}\,,
 $$
 see \cite{Naz11}, p.91, and \cite{KM15}. Excluding $s_3$ using the first relation we write the second as
 $-2(k_1-k)\cdot(k_2-k)=0$.  Or $\ -2 x\cdot y=0$, if we denote $x=k_1-k$, $y=k_2-k$. That is, $K_k$
 is given by an integral over the set $\Sigma_*\subset \R^{2d}$ as in \eqref{p22}. 
 In a work in progress (see
 \cite{Oberv}) we make an attempt to derive rigorously a wave 
 kinetic equation for NLS with added small dissipation
 and small random force (see \cite{KM15, Oberv} for a discussion of this model). On this way nonlinearities of
 the form \eqref{heur} appear naturally as limits for $\nu\to0$ of certain integrals of the form \eqref{I_s},
 where, again, $x=k_1-k$, $y=k_2-k$. \footnote{So the integrand $F_k$ depends on the parameter $k\in\R^d$. 
 This dependence should be controlled, which can be done  with some extra efforts.}
 We strongly believe that more asymptotical expansions of integrals, similar to \eqref{I_s}, will appear 
 when more works on rigorous justification of physical methods to treat nonlinear waves will come out. 
 
 Proof of Theotem \ref{t_singint}, given below in Sections \ref{s_71}--\ref{s4}, is rather general
and applies to other integrals with singular divisors.  Some of these applications  are discussed 
in Section~\ref{s_other_integrals}. 
 \medskip
 
\noindent 
{\bf Notation.}
By $\chi_A$ we denote the characteristic function of a set $A$. 
For an integral $I=\int_{\R^{2d}} f(z)\,dz$  and a submanifold $M\subset\R^{2d}$, dim$\,M=m\le 2d$,
compact or not (if $m=2d$, then $M$ is an open domain in $\R^{2d}$) we write
$$
\lan I,M\ran =  \int_Mf(z)\,d_M(z), 
$$
where $d_M(z)$ is the volume--element on $M$, induced from $\R^{2d}$. 
\smallskip

\noindent 
{\bf Acknowledgments.} We acknowledge the support from the Centre National de la Recherche Scientifique (France)
 through the grant PRC CNRS/RFBR 2017-2019 No~1556 
 ``Multi-dimensional semi-classical problems of condensed matter physics and quantum dynamics'',
 and thank Johannes~Sj\"ostrand for explaining the way to estimate singular integrals 
  \eqref{another_integral}, presented in Appendix to this work.

\section
{Integral over the vicinity of $0$.}\label{s_71}

For $ 0<\delta\le1$ consider the domain 
$$
K_\delta = \{|x|\le\delta, |y|\le\delta\}\subset \R^d\times \R^d \,,
$$
and the integral 
\be\label{int_delta}
    \int_{K_\delta} \frac{ |F(x,y)|\,dx\,dy}{(x\cdot y)^2 +( \nu \Gamma(x,y))^2}\,.
\ee
Obviously, everywhere in $K_\delta$,
$
|F(x,y)|\le C_1$ and $  \Gamma(x,y) \ge C.
$
So the  integral  is bounded by $ C_1  \bar I_\nu(\delta)$, where 
$$
 {\bar I_\nu}(\delta)=    \int_{|x|\le\delta}  \int_{|y|\le\delta}  \frac{dx\,dy}{ (x\cdot y)^2 + (C\nu )^2}\,.
$$
We write ${\bar I_\nu}(\delta)$ as
$$
{\bar I_\nu}(\delta) = \int_{|x|\le\delta} J_x\,dx\,,\qquad 
J_x = \int_{ |y|\le \delta} \frac{dy}{ (x\cdot y)^2 +(C\nu)^2}\,.
$$
Let us 
 introduce in the $y$--space a coordinate system $(y_1, \dots,y_d)$ 
 with the first basis vector $e_1 = x/r$, where $r=|x|$. Since  the volume of the layer, lying in 
 the ball $\{  |y|\le \delta\}$ above  an infinitesimal segment $[{y_1}, {y_1}+d{y_1}]$  is $\le C_d \delta^{d-1} d{y_1}$
and since $(x\cdot y) = r{y_1}$, then 
\begin{equation*}
\begin{split}
J_x   
 \le C_d  r^{-2} \int_0^\delta d{y_1}\, \frac{ \delta^{d-1} }{{y_1}^2+(C\nu/r)^2}
=C_d \delta^{d-1} \frac{\tan^{-1}(r\delta/C\nu)}{Cr\nu}
\le \frac{\pi}2 C_d\, \frac{\delta^{d-1}}{  Cr\nu}\,.
\end{split}
\end{equation*}
So 
$$
{\bar I_\nu}(\delta)  = \int_{|x|\le\delta} J_x dx \le   C_d  \frac{\delta^{d-1}}{C\nu}  \int_0^\delta r^{d-2}\,dr 
\le  C'_d{\delta^{2d-2}}{\nu^{-1} } \,.
$$
Thus we have proved

\begin{lemma}\label{l_nearsing}
The integral  \eqref{int_delta} 
is bounded by 
$
  {C} {\nu^{-1} \delta^{2d-2}} \,. 
$
\end{lemma}

Now we pass to the global study of the integral \eqref{I_s} and begin with studying 
the geometry of the manifold $\Sigma_*$ and its vicinity in $\R^{2d}$.

\section{The manifold $\Sigma_*$ and its vicinity.} \label{s_3}
 The set $\Sigma_*=\Sigma\setminus (0,0)$ is a smooth submanifold of $\R^{2d}$  of dimension $2d-1$. 
Let $\xi\in\R^{2d-1}$ be a local coordinate on $\Sigma_*$ with the coordinate mapping
$\xi\mapsto (x_\xi,y_\xi)=z_\xi\in \Sigma_*$. Abusing notation we 
write $|\xi|= |(x_\xi,y_\xi)|$. The vector 
$
N(\xi)= (y_\xi, x_\xi)
$
is a normal to $\Sigma_*$ at $\xi$ of  length $|\xi|$, and
\be\label{orth}
 N(\xi)\cdot (x_\xi, y_\xi)  = 2 \, x_\xi\cdot y_\xi =0\,.
\ee
For any $0\le R_1<R_2$ we denote 
\be\label{nota}
\begin{split}
&S^{R_1} = \{z\in\R^{2d}: |z|=R_1\}\,,\quad  \Sigma^{R_1} = \Sigma\cap S^{R_1} \,,
\\
& S^{R_2}_{R_1} = \{z: R_1< |z|<R_2\}\,,\quad \Sigma^{R_2}_{R_1}  = \Sigma\cap S^{R_2}_{R_1}\,,
\end{split}
\ee
and for  $t>0$  denote by $D_t$ the dilation operator
$$
D_t:  \R^{2d}\to \R^{2d},\quad z \mapsto tz\,.
$$
It preserves $\Sigma_*$, and for any $ \xi\in\Sigma_*$ we denote by $t\xi$
the point $D_t(x_\xi, y_\xi).$

\begin{lemma}\label{l_p1}
1)
There exists $\theta_0^*\in(0, 1]$
such that for any $0<\theta_0\le \theta_0^*$ a
 suitable \nbh\  $\Sv=\Sv(\theta_0)$ of $\Sigma_*$ in $\R^{2d}\setminus\{0\}$ may be uniquely parametrised 
 as
\be\label{par}
\Sv=
\{ \pi(\xi,\theta): 
\xi\in\Sigma_*,\; |\theta|<\theta_0\}\,,
\ee
where $\pi(\xi,\theta) = (x_\xi,y_\xi) +\theta N_\xi = (x_\xi,y_\xi) + \theta(y_\xi, x_\xi) $.\\
2) For any vector $\pi = \pi(\xi, \theta)\in\Sv$  its length equals  
\be\label{length}
|\pi| =|\xi|\sqrt{1+\theta^{2}}.
\ee
The distance from $\pi$ to
$\Sigma$ equals $|\xi| |\theta|$, and the shortest path from $\pi$ to $\Sigma$  is the segment 
$ [\xi,\pi]= \{\pi(\xi, t\theta): 0\le t\le 1\}=:S$. \\
3) If $z=(x,y)\in S^R$ is such that $\dist (z, \Sigma)\le \frac12 R\theta_0$, then $z=\pi(\xi,\theta)\in\Sv$,
where $|\theta|<\theta_0$ and 
$|\xi|\le R\le |\xi|\sqrt{1+\theta_0^2}$. \\
4) If $\pi(\xi,\theta)\in\Sv$, then 
 \be\label{p0}
 \omega \big( \pi(\xi,\theta)\big) =|\xi|^2\theta.
  \ee
  5) If $(x,y)\in S^R\cap (\Sv)^c$, then $|x\cdot y| \ge cR^2$ for some $c=c(\theta_0)>0$. 
\end{lemma}

The 
coordinates \eqref{par} are known as the normal coordinates, and their existence  follows easily
from the implicit function theorem. The assertion 1) is a bit more precise than the general result since it 
specifies the size of the \nbh\ $\Sv$.

\begin{proof}
1) Fix any  positive $\kappa<1$. Then for $\theta_0^*$  small enough  it is well known 
that the points $\pi(\xi,\theta)$ with $\xi\in\Sigma^{1+\kappa}_{1-\kappa}$ and
$|\theta|<\theta_0 \le\theta_0^*$ 
 form a \nbh\ of 
$\Sigma^{1+\kappa}_{1-\kappa}$ in $\R^{2d}$ and parametrise it in a unique and smooth way.
Besides, any point 
$\pi'\in\R^{2d}$ such that dist$\, (\pi', \Sigma^1)\le \frac12\theta_0$,  may 
be represented as 
\be\label{p3}
\pi' = \pi(\xi', \theta'), \quad \xi'\in\Sigma^{1+\kappa}_{1-\kappa},\;\; |\theta'|<\theta_0\,,
\ee
and
\be\label{p39}
\pi(\xi_1, \theta_1) =\pi(\xi_2, \theta_2),\; \xi_1, \xi_2\in \Sigma^{1+\kappa}_{1-\kappa} \ 
\Rightarrow\  |\theta_1|,  |\theta_2| \ge 2\theta_0^*\,.
\ee
We may assume that 
$
\theta_0^*<\tfrac12\kappa. 
$
The mapping $D_t$ sends $\Sigma^{1+\kappa}_{1-\kappa}$ to  $\Sigma^{t+t\kappa}_{t-t\kappa}$ and sends 
$\pi\big( \Sigma^{1+\kappa}_{1-\kappa}\times(-\theta_0,\theta_0)\big)$ to 
$\pi\big( \Sigma^{t+t\kappa}_{t-t\kappa}\times(-\theta_0,\theta_0)\big)$. This implies that the set $\Sv$, defined as a collection of all points $\pi(\xi,\theta)$ as
in \eqref{par}, makes a \nbh\ of $\Sigma_*$. To prove that the parametrisation is unique assume that it is not. Then there exist
$t_1>t_2>0$, 
$\theta_1, \theta_2\in(-\theta_0, \theta_0)$ and $\xi_1\in\Sigma^{t_1}, \xi_2\in\Sigma^{t_2}$ such that 
$\pi(\xi_1,\theta_1)=\pi(\xi_2,\theta_2)$. So $\pi_1=\pi_2$, where 
$$
\pi_1= \pi(t_1^{-1} \xi_1,\theta_1),\quad \pi_2=  \pi(t_1^{-1} \xi_2,\theta_2),
$$
and $1=|t_1^{-1} \xi_1|> |t_1^{-1} \xi_2|$. Let us write $|t_1^{-1} \xi_2|$ as $1-\kappa', \kappa'>0$. If $\kappa'<\kappa$, then
 $(\xi_1,\theta_1)=(\xi_2,\theta_2)$ by what was said above. If $\kappa'>\kappa$, then 
 $|t_1^{-1}\xi_1 - t_1^{-1}\xi_2| \ge |t_1^{-1}\xi_1| - |t_1^{-1}\xi_2| \ge
 \kappa$. Since 
  $$
 |\pi_1-t_1^{-1}\xi_1|=\theta_1,\qquad  |\pi_2-t_2^{-1}\xi_2|=|\theta_2 N_{t_1^{-1}\xi_2}|\le
 \theta_2\,,
 $$
 then $|\pi_1-\pi_2|\ge \kappa-\theta_1-\theta_2\ge \kappa-2\theta_0$. Decreasing $\theta_0^*$ if needed, we achieve that  $\kappa>2\theta_0$, so $|\pi_1-\pi_2|>0$. Contradiction. 
 
 2) The first assertion holds since by \eqref{orth} the vector $N_\xi$  is orthogonal to $(x_\xi,y_\xi)$  and since its
 norm equals $|\xi|$. 
 The second assertion holds since the segment $S$ is a geodesic from $\pi$ to $\Sigma_*$, orthogonal to $\Sigma_*$.
 Any other geodesic from $\pi$ to $\Sigma_*$ must be a segment $S'= [\pi,\xi']$, $\xi'\in\Sigma_*$,
 orthogonal to $\Sigma_*$. 
 It is longer than $|\theta| |\xi|$.
 To prove this, by scaling (i.e. by applying a delation operator), we reduce the problem to the case $|\xi|=1$. Now,
 if $\xi'\in \Sigma^{1+\kappa}_{1-\kappa}$, then $\pi(\xi',\theta') = \pi =\pi(\xi,\theta)$ for some real 
 number $\theta'$. So by \eqref{p39}, $|\theta| \ge 2 |\theta^*_0|$, which is a contradiction. 
 While if  $\xi'\not\in \Sigma^{1+\kappa}_{1-\kappa}$, then the distance from $\pi$ to $\xi'$ is bigger than $\kappa - \theta_0>\theta_0$. Indeed, if $|\xi'|\ge 1+\kappa$, then the distance is bigger than
 $
 1+\kappa -|\pi| \ge 1+\kappa-1 -\theta_0=\kappa-\theta_0. 
 $
The case $|\xi'|\le 1-\kappa$ is similar.

 3) If $R=1$, then the assertion follows from \eqref{p3} and \eqref{length}. 
 If $R\ne1$, we apply the operator $D_{R^{-1}}$ and use the result with $R=1$.
 
 4) Follows immediately from \eqref{orth}. 
 
 5) If $R=1$, then the  assertion with some $c>0$ follows from
 the compactness of $S^1\cap (\Sv)^c$. If $R\ne1$, then again we apply   $D_{R^{-1}}$ and use the result with $R=1$.
 \end{proof}
 \smallskip
 
 Let as fix any $0<\theta_0\le\theta_0^*$, and consider the manifold  $\Sv=\Sv(\theta_0)$. Below we provide it with some additional 
 structures and during the corresponding constructions decrease $\theta_0^*$, if needed. 
 Consider the set $\Sigma^1$. It equals 
 $$
 \Sigma^1= \{(x,y): x\cdot y=0, x^2+y^2=1\}\,.
 $$
 Since the differentials of the two relations, defining $\Sigma^1$, are independent on  $\Sigma^1$, 
  then this set  is a
 smooth compact  submanifold of $\R^{2d}$ of codimension 2. Let us cover it by some finite system of charts
   $\cN_1,\dots, \cN_{\tilde n}$, $\cN_j =\{\eta^j = (\eta_1^j,\dots,\eta^j_{2d-2})\}$.  Denote by $m(d\eta)$ the volume element
   on $\Sigma^1$, induced from $\R^{2d}$, and denote the  coordinate maps as 
   $\cN_j\ni \eta^j\to(x_{\eta^j},y_{\eta^j})\in\Sigma^1$. We will write points of $\Sigma^1$ both as $\eta$ and $(x_{\eta},y_{\eta})$.
   
   The mapping 
   $$
   \Sod\times \R^+ \to \Sigma_*,\quad ( (x_{\eta},y_{\eta}), t)\to D_t (x_{\eta},y_{\eta}),
   $$
   is 1-1  and is a local diffeomorphism; so this is a global diffeomorphism. Accordingly,
    we can cover $\Sigma_*$ by the $\tilde n$
   charts $\cN_j\times \R_+$, with the coordinate maps
   $$
   (\eta^j, t) \mapsto D_t(x_{\eta^j},y_{\eta^j}),\quad \eta^j\in\cN_j,\;\; t>0\,,
   $$
   and can apply Lemma \ref{l_p1}, taking $(\eta,t)$     for the coordinates  $\xi$. 
   In these coordinates the volume element on $\Sigma^t$ is $t^{2d-2}m(d\eta)$.  Since $\p/\p t \in T_{\eta,t}\Sigma_*$ is 
   a vector of unit length, perpendicular to $\Sigma^t$,\footnote{as $\p/\p t\perp S^t$ and $S^t
   \supset \Sigma^t$.
   }
   then the volume element     on $\Sigma_*$ is 
   \be\label{vol_on_*}
   d_{\Sigma_*}=  t^{2d-2}m(d\eta)\,dt\,.
   \ee
   
   The coordinates $(\eta,t,\theta)$ with $\eta \in\cN_j, t>0, |\theta|<\theta_0$, where $1\le j\le\tilde n$,  make  coordinate systems on the open set 
   $\Sv$. Since the vectors $\p/\p t$ and $t^{-1} \p/\p \theta$ form an orthonormal base of the orthogonal complement in  $\R^{2d}$
 to $T_{(\eta,t,0)} \Sigma^t$,\footnote{Since the vector $\p/\p t$ is perpendicular to $\Sigma^t$ and 
 lies in $T_{(\eta, t,0)}\Sigma_*$, and $\p/\p\theta$ is proportional to the vector $N_{(\eta, t,0)}$, normal to 
 $\Sigma_*$ at ${(\eta, t,0)}$.
 }
 then in   $\Sv$ the volume element $dx\,dy$ may be written as
 \be\label{p4}
 dx\,dy =t^{2d-1} \mu(\eta,t,\theta) m(d\eta)dt\,d\theta\,, \quad \text{where}\;\;
 \mu(\eta,t,0)=1\,.
 \ee
 For $r>0$
 the transformation $D_r$ multiplies the form in the l.h.s. by $r^{2d}$, preserves $d\eta$ and $d\theta$, and multiplies
  $dt$ by $r$. Hence, $\mu$ does not depend on $t$, and we have got 
  
  \begin{lemma}\label{l_p2}
  The coordinates
  \be\label{chj}
  (\eta^j, t,\theta), \;\;\text{where}\;\; \eta^j\in\cN_j,
   \; t>0, \; |\theta|<\theta_0\,,
  \ee
and   $1\le j \le\tilde n$,   define on $\Sv$ coordinate systems, jointly covering $\Sv$. In these coordinates 
 the dilations $D_r$, $r>0$,  reed as
$$
D_r: (\eta, t,\theta) \mapsto (\eta, rt, \theta)\,,
$$
and the volume element has the form \eqref{p4}, where $\mu$ does not depend on $t$. 
  \end{lemma}
  
  Besides, since at a point $z=(x,y)=\pi(\xi,\theta)\in\Sv$ we have 
  $(\p/\p\theta) = \nabla_z\cdot (y,x)$, then in view of \eqref{F_1}, \eqref{Ga_1}
  \be\label{new_est}
  \big| \frac{\p^k}{\p \theta^k} F(\eta, t, \theta) \big|  \le K'(1+t)^{-M},\;\; 
  \big| \frac{\p^k}{\p \theta^k} \Gamma(\eta, t, \theta) \big| \le K'(1+t)^{r_*}
  \ee
  for $0\le k\le2$ and for all $(\eta, t, \theta)$. 
  
  For $0\le R_1<R_2$ we  denote
  $$
  \big(\Sv\big)^{R_2}_{R_1} =  \pi \big(\Sigma^{R_2}_{R_1} \times(-\theta_0,\theta_0)\big)\,.
  $$
  In a chart \eqref{chj} this domain is $\{(\eta^j, t,\theta): \eta^j\in\cN_j, \ R_1<t< R_2,\  |\theta|<\theta_0\}$.

  \section{Global study of the integral \eqref{I_s}} \label{s4}
    
 \subsection{Desintegration of $I_\nu$}
 Using \eqref{p4}, for any $0\le R_1<R_2$ we write the integral
  $\lan  I_\nu, \big(\Sv\big)^{R_2}_{R_1}\ran$ as
 \be\label{3.9}
 \begin{split}
 \int_{\Sod} m(d\eta) \int_{R_1}^{R_2} dt\,  t^{2d-1}   \int_{-\theta_0}^{\theta_0} d\theta
 \frac{F(\eta, t,\theta) \mu(\eta,\theta)}{(x\cdot y)^2 +( \nu \Gamma(\eta,t,\theta))^2}\\
=  \int_{\Sod} m(d\eta) \int_{R_1}^{R_2} dt \, t^{2d-1} J_\nu(\eta,t)\,,
 \end{split}
 \ee
 where by \eqref{p0} 
 $$
 J_\nu(\eta,t) = \int_{-\theta_0}^{\theta_0} d\theta 
 \frac{F(\eta, t,\theta) \mu(\eta,\theta)}{t^4\theta^2 +( \nu \Gamma(\eta,t,\theta))^2}\,.
 $$
  To study $ J_\nu(\eta,t)=:J_\nu$ we write $\Gamma$ as
  $$
  \Gamma(\eta,t,\theta) = h_{\eta, t}(\theta) \Gamma(\eta, t,0)\,. 
  $$
  The function $ h(\theta):= h_{\eta, t}(\theta) $ is $C^2$-smooth, and in view of 
   \eqref{new_est} 
   and \eqref{Ga_1} it  satisfies 
  \be\label{p11}
  |h(\theta)|\ge C_0^{-1},\quad 
  \Big| \frac{\p^k}{\p \theta^k} h(\theta) \Big| \le C_k\quad \forall\, \eta, t, \theta, \; 0\le k\le2  \,.
  \ee
  Denoting $$\eps=  \nu t^{-2} \Gamma(\eta, t,0),$$ we write $J_\nu$ as 
  $$
  J_\nu = t^{-4} \int_{-\theta_0}^{\theta_0} 
  \frac{F(\eta,t,\theta)\mu(\eta,\theta) h^{-2}(\theta)\,d\theta}{\theta^2 h^{-2}(\theta) +\eps^2}\,.
  $$
  Since $h(0)=1$, then in view of \eqref{p11} the mapping
  $$
  f= f_{\eta,t}: [-\theta_0, \theta_0]\ni \theta \mapsto \thb = \theta  / h(\theta) 
  $$
  is a $C^2$--diffeomorphism on its image 
  such that $f(0)=0, f'(0)=1$ and 
   the $C^2$-norms of $f$ and  $f^{-1}$  are bounded by a constant, independent from $\eta, t$
  (to achieve that,   if needed, we decrease $\theta_0^*$). Denote 
  $$
  \theta_0^+ = f(\theta_0),\quad 
  \theta_0^-=-f(-\theta_0),\quad\thb_0=\min(\theta_0^+, \theta_0^-)\,.
  $$
  Then 
  $
  2^{-1} \theta_0 \le \theta_0^\pm \le 2 \theta_0
  $ 
 if  $\theta_0^*$ is small, and  
  $$
  J_\nu = t^{-4} \int_{\theta_0^-}^{\theta_0^+} 
  \frac{F(\eta,t,\theta)\mu(\eta,\theta) h^{-2}(\theta) (f^{-1}(\theta))' \,d\thb}{\bar\theta^2  +\eps^2}\,.
  $$
 Denote the nominator of the integrand as $\Phi(\eta, t,\thb)$.  This is a $C^2$--smooth function, and 
 by \eqref{new_est} and \eqref{p11} it satisfies
 $$
 | \frac{\p^k}{\p\theta^k} \Phi| \le C(1+t)^{-M}\quad \text{for}\quad 0\le k\le 2\,.
 $$
   Moreover, since $h(0)=1$ and $(f^{-1}(0))'= f'(0)=1$, then in view of \eqref{p4} we have that 
\be\label{p13}
\Phi(\eta, t, 0) = F(\eta, t, 0)\,.
\ee

Consider the interval 
$\Delta_{\eta, t} = f_{\eta,t}^{-1} (-\thb_0, \thb_0)$. Then
$$
(-\theta_0/2, \theta_0/2) \subset \Delta_{\eta, t} \subset (-\theta_0, \theta_0)
$$
for all $\eta$ and $ t$.  Now we modify the \nbh  \ $\Sigma^{nbh}(\theta_0)$ 
to
$$
\Svm= \Svm(\theta_0) = \{ \pi(\eta, t,\theta): \eta\in \Sigma^1,\  t>0,\  \theta\in \Delta_{\eta,t}\}\,.
$$
Then
\be\label{Smod}
\Sv (\tfrac12 \theta_0) \subset \Svm(\theta_0) \subset \Sv(\theta_0)\,.
\ee
The modified analogy  $J_\nu^m$ of the 
integral  $J_\nu$ has the same form as $J_\nu$, but the domain of integrating becomes not
$(-\theta_0, \theta_0)$, but $\Delta_{\eta, t} $. Then 
 $$
  J_\nu^m = t^{-4} \int_{-\thb_0}^{\thb_0} 
  \frac{\Phi(\eta,t,\theta) \,d\thb}{\bar\theta^2  +\eps^2}\,.
  $$


To estimate  $J_\nu^m$, 
consider first the integral $J_\nu^{0m}$,
 obtained from $J_\nu^{m}$ by frozening $\Phi$ at $\thb=0$:
$$
 J_\nu^{0m}= t^{-4}  \int _{-\thb_0}^{\thb_0}\frac{ \Phi(\eta,t,0)\,d\,\thb}{\thb^2+\eps^2}
  = 2t^{-4} F(\eta,t,0) \eps^{-1} 
  \tan^{-1}\frac{\thb_0}{\eps}
  $$
  (we use \eqref{p13}). From here 
  \be\label{p_triv}
  | J_\nu^{0m}| \le \pi \eps^{-1}t^{-4} |F(\eta,t,0)|\,.
  \ee
  As $0<\frac{\pi}2 -\tan^{-1}\frac1{\bar\eps}<\bar\eps$ for $0<\bar\eps\le\frac12$, then also 
  \be\label{p17}
  0<\pi \nu^{-1} t^{-2} 
  (F/\Gamma)\mid_{\theta=0}
  - J_\nu^{0m} <\frac2{\thb_0} t^{-4} F(\eta,t,0)\,,
  \ee
  if
  \be\label{p18}
  \nu t^{-2} \Gamma(\eta,t,0)\le \tfrac12 \thb_0.
  \ee

  Now we estimate the difference between $J_\nu^{m}$ and $J_\nu^{0m}$. We have:
  $$
  J_\nu^{m}-J_\nu^{0m} = t^{-4} \int_{-\thb_0}^{\thb_0} \frac{\Phi(\eta,t,\thb)-\Phi(\eta,t,0)}{\thb^2+\eps^2}d\thb\,.
  $$
  Since 
  each  $C^k$--norm   of   $\Phi$ , $k\le 2$, 
   is bounded by $C(1+t)^{-M}$, then
  $$
  \Phi(\eta,t,\thb)-\Phi(\eta,t,0) = A(\eta,t)\thb +B(\eta,t,\thb)\thb^2\,,
  $$
  where $|A|, |B| \le C(1+t)^{-M}$. From here
  $$
  |J_\nu^{m} - J_\nu^{0m}|\le C_1(1+t)^{-M} t^{-4} \int_0^{\thb_0} \frac{\thb^2\,d\thb}{\thb^2+\eps^2}\le 
  C_1(1+t)^{-M}t^{-4}\thb_0\,.
  $$
  Denote 
  \be\label{p20}
  \cJ_\nu(\eta,t)= \pi t^{-2} \big( F\Gamma^{-1}\big)(\eta,t,0)\,.
  \ee
  Then, jointly with \eqref{p17}, the last estimate tell us that 
  \be\label{p19}
  |J_\nu^{m} - \nu^{-1}\cJ_\nu(\eta,t)|\le C(1+t)^{-M}t^{-4}\thb_0^{-1}\quad\text{if \eqref{p18} holds}.
  \ee
  
  \subsection{Proof of Theorem \ref{t_singint}} \label{s_76} We have to distinguish the cases  $r_*\le2$ and
  $r_*>2$. 
  
  \noindent
  {\bf Let $r_*\le2$.} Then by  \eqref{Ga_1} assumption \eqref{p18} holds if $ t\ge C^{-1}\sqrt\nu$ (where 
  $C$ depends on $\thb_0$). Integrating $J^m_\nu(\eta,t)$ and $ \nu^{-1}\cJ_\nu(\eta,t)$ with respect to the 
  measure $t^{2d-1} m(d\eta)\,dt$ and using \eqref{p19}, \eqref{3.9}  and  \eqref{hr} we get that 
    \be\label{p41}
  \begin{split}
 & |\lan I_\nu, (\Svm)^\infty_{C^{-1}\sqrt\nu}\ran - \nu^{-1} 
   \int_{\Sigma^1} m(d\eta) \int_{C^{-1}\sqrt\nu}^\infty dt\,  t^{2d-1} \cJ_\nu(\eta, t)| \\
 & \le C \int_{\Sigma^1} m(d\eta) \int_{C^{-1}\sqrt\nu}^\infty dt\,  t^{2d-1}t^{-4}  (1+t)^{-M}
  \le C  \chi_d(\nu)
  \end{split}
  \ee
  (for the quantity $\chi_d(\nu)$   see \eqref{chi_d}). 
   In view of \eqref{p20} and Lemma \ref{l_nearsing} with $\delta=2C^{-1}\sqrt\nu$,
  \be\label{p42}
  \begin{split}
 & |\lan I_\nu, (\Svm)^{C^{-1}\sqrt\nu}_0\ran - \nu^{-1}\int_{\Sigma^1}m(d\eta) \int_0^{C^{-1}\sqrt\nu}
  dt\, t^{2d-1} \cJ_\nu(\eta,t) |\\
 &\le C\nu^{-1} \nu^{d-1} + C\nu^{-1} \int_0^{C^{-1}\sqrt\nu} t^{2d-1} t^{-2}\,dt\le C_1\,
  \end{split}
  \ee
  as $d\ge2$.
  Next, by \eqref{p20} and \eqref{hr} 
  \be\label{p43}
  \int_{\Sigma^1}m(d\eta) \int_0^\infty dt\, t^{2d-1}| \cJ_\nu(\eta,t)| \le C \int_0^\infty t^{2d-1-2}(1+t)^{-M-r_*}\le C_1\,,
  \ee
  and by \eqref{vol_on_*}
  \be\label{p44}
   \begin{split}
  \int_{\Sigma^1}m(d\eta) \int_0^\infty  dt\, t^{2d-1} \cJ_\nu =& \pi 
  \int_{\Sigma^1}m(d\eta) \int_0^\infty dt\, t^{2d-3}(F/\Gamma)\mid_{\theta=0}\\
  =
  &\pi\int_{\Sigma_*} |z|^{-1} (F/\Gamma)(z) d_{\Sigma_*}z\,.
 \end{split}
  \ee
  
  This gives us asymptotic description as $\nu\to0$ of the integral \eqref{I_s}, calculated
  over the vicinity $\Svm$ of $\Sigma_*$. It remains to estimate the integral over the complement 
  to $\Svm$. But this is easy: by \eqref{Smod}, 
 \begin{equation*}
  \begin{split}
| \lan I_\nu, \R^{2d}\setminus \Svm\ran|
&\le |
\lan I_\nu,   \{|(x,y)|\le 2 \nu\} \ran |\\
&+  C_d \Big| \int_{\nu}^\infty dr\, r^{2d-1} 
\int_{S^r\setminus \Sv( \theta_0/2)} \frac{F(x,y)\, d_{S^r}}{ (x\cdot y)^2 +(\nu\Gamma((x,y))^2}\Big|\,.
  \end{split}
  \end{equation*}
  By item 5) of Lemma~\ref{l_p1} the divisor of the integrand  is $\ge C^{-2} r^4$. Due to this and
  \eqref{F_1}, the second term in the r.h.s.    is bounded by
  $$
    C \int_{\nu}^\infty (1+r)^{-M} r^{2d-5}\,dr
  \le C_1   \chi_d(\nu)\,.
  $$
  This estimate     and Lemma~\ref{l_nearsing}   with $\delta=2\nu$  imply that 
  \be\label{p45}
| \lan  I_\nu, \R^{2d}\setminus \Svm\ran | \le C   \chi_d(\nu)\,.
  \ee
  
  Now relations \eqref{p41}, \eqref{p42}, \eqref{p44}, \eqref{p45} imply \eqref{p22},
  while \eqref{vol_on_*} and \eqref{p43} imply that the integral in \eqref{p22} converges 
  absolutely. 
  \smallskip

  \noindent
  {\bf Let $r_*>2$.}  Then condition \eqref{p17} holds if 
  $$
  C_\beta \nu^{-\beta} \ge t \ge C^{-1} \sqrt\nu\,,\qquad
  \beta=\frac1{r_*-2}\,.
  $$
  Accordingly, the term   in the l.h.s. of \eqref{p41} should be split in two. The first corresponds to
  the integrating from $C^{-1}\sqrt\nu$ to $C_\beta\nu^{-\beta}$ and estimates exactly as before. The 
  second is 
  \be\label{p02}
   |\lan I_\nu, (\Svm)^\infty_{C_\beta \nu^{-\beta}}\ran - \nu^{-1} 
   \int_{\Sigma^1} m(d\eta) \int_{C_\beta \nu^{-\beta}}^\infty dt\,  t^{2d-1} \cJ_\nu(\eta, t)|\,.
  \ee
  To bound it  we estimate the norm of the difference of the two integrals via the sum of their norms. In
  view of \eqref{p_triv} and \eqref{p20}  both of them are bounded by 
  $$
  C\nu^{-1} \int_{\Sigma_1}m(d\eta)\int_{C_\beta \nu^{-\beta}}^\infty dt\,  t^{2d-3}
  (F/\Gamma)(\eta, t, 0)\,.
  $$
  So
  $$
  \eqref{p02}\,  \le \ C_\beta \nu^{-1+\beta(M+r_*+2-2d)}\le C_\beta
  $$
  since $M>2d-4$.
  
  Adding this relation to   \eqref{p41}, applied to the 
  integrating from $C^{-1}\sqrt\nu$ to $C_\beta\nu^{-\beta}$, and -- as before -- using this jointly with 
   \eqref{p42}, \eqref{p44}, \eqref{p45}, we again get \eqref{p22} 
   (while the absolute convergence of the integral  still follows from  \eqref{p43}).
   \qed

\subsection{Proof of Proposition \ref{prop}}   
  For any  $R>r>0$ let us denote by $B^R$ and $B_r^R$
  the ball $\{|z|\le R\}$ and the set $\{ r\le  |z|\le R\}$,
  and 
   consider the  measure 
  $\mu^R_r = \chi_{B^R_r} \, (z) |z|^{-1} \delta_{\Sigma_*}$.
   This is a well defined  Borel measure on  $\Sigma_*$ and on 
   $\R^{2d}$. As $\eps\to0$, the measures $\mu^R_\eps$ weakly converge to a limit. This is the restriction of
   the   measure
    $|z|^{-1} \delta_{\Sigma_*}$  to $B^R$, and its further  restriction to $B^R_r$ equals $\mu^R_r$.
    So $ |z|^{-1} \delta_{\Sigma_*}$ is a $\sigma$--additive Borel measure on $\R^{2d}$, and it has no  atoms 
     outside the origin. Let us abbreviate  $ |z|^{-1} \delta_{\Sigma_*}=\mu.$
  By \eqref{vol_on_*}, 
   for any $0<r\le R$ and  $\eps\le r$, 
  $$
  \int_{B^R_r} \,d\, \mu
  = \int_{B^R_r} \,d\, \mu^R_\eps
   = \int_r^R t^{2d-2}t^{-1} \, dt = \frac1{2d-2}(R^{2d-2} -r^{2d-2}).
 $$
  From here and the weak convergence of the measures $\mu^R_\eps$ to $\mu\mid_{B^R}$ we get that 
  $\mu\{|z|<\rho\} \le \rho^{2d-2}/(2d-2)$ if $\rho<R$, so $\mu$  has no atom in the origin and  is atomless. Next,  
   for any function $f\in C_m(\R^{2d})$ we have
  \begin{equation*}
  \begin{split}
 & \int  f(z)\, \mu(dz) \le 
  |f|_m  \sum_{R=0}^\infty \int_{B^{R+1}_R} \lan z\ran^{-m} \, \mu(dz)\\
 & \le
  C_1 |f|_m  \sum_{R=0}^\infty \lan R\ran^{-m}  
  \frac{(R+1)^{2d-2} -R^{2d-2}}{2d-2}\\
 & \le 
  C_2 |f|_m  \sum_{R=0}^\infty \frac{(R+1)^{2d-1} }{\lan R\ran^{-m}}
    = C_3 |f|_m,
    \end{split}
   \end{equation*}
  if $m>2d-2$. This proves the proposition.

\section{Other integrals}\label{s_other_integrals}
The geometrical approach to treat integrals \eqref{I_s}, developed above, 
applies to various modifications of these integrals. Below we briefly discuss three more examples.

\subsection{Integrals \eqref{I_s} with $d=1$} \label{s_d1}
The restriction $d\ge2$ was imposed in the previous sections since in the one-dimensional case
some integrals, involved in the construction, strongly diverge at the locus of the quadric $\Sigma$. 
This problem disappears if the function $F$ vanishes near the locus. Indeed, consider 
 $$
 I'_\nu =  \int_{\R^2} \frac{F(x,y)  }{x^2y^2 +(\nu\G(x,y))^2}\, dxdy\,,
 $$
 where $F\in C^2_0(\R^2)$ vanishes near the origin. The quadric $\Sigma' = \{xy=0\}\subset\R^2$ is one dimensional, 
 has a singularity at the origin, and its smooth part ${\Sigma'}^{*} = \Sigma'\setminus {0}$ has four connected components. 
 Consider one of them:
 $
 \cC_1=\{(x,y): y=0, x>0\}.
 $
 Now the coordinate 
 $\xi$  is a point in $\R_+$ with $(x_\xi, y_\xi)= (\xi,0)$ and  with the normal $N(\xi)=(0,\xi)$, 
 the set $\Sigma_1\cap \cC_1$ is the single  point $(1,0)$
 and the coordinate $(\eta,t,\theta)$ in the vicinity of $\cC_1$ degenerates to $(t,\theta)$, $t>0$, $|\theta|<\theta_0$, with the coordinate-map
 $(t,\theta) \mapsto (t, t\theta)$. The relations \eqref{vol_on_*} and \eqref{p4} are now 
 obvious, and the integral \eqref{int_delta} vanishes if $\delta>0$ is 
sufficiently small. Interpreting  $z=(x,y)$ as a complex number, we write the assertion of Theorem \ref{t_singint} as
$$
\big| J'_\nu  - \pi \nu^{-1} 
 \int_{\Sigma'} \frac{F(z)}{|z| \, \G(z)}\,d  z\big| \le C\,,
$$
where  the integral is a contour integral in the complex plane.

\subsection{Integrals of quotients with divisors, linear in $\omega$.}
Consider
\be\label{another_integral}
I'_\nu = \int_{\R^{2d}} \frac{F(x,y)\,dxdy}{x\cdot y +i\nu \Gamma(x,y)}\,.
\ee
Now there is no need to separate the integral over the vicinity of the origin, and we just  split $I'_\nu$
to an integral over $\Svm$ and over its complement.

To calculate $\lan I'_\nu, \Svm\ran$  we observe that 
an analogy of  $J_\nu^{0m}$ is the integral 
$$
 \int_{-\bar\theta_0}^{\bar\theta_0} 
\frac{F(\eta, t,0)\,d\bar\theta}{t^2 \bar\theta+ i\nu\Gamma(\eta,t,0)}
=t^{-2} F(\eta, t,0) \ln\frac{\theta_0^+ + i\nu t^{-2}\Gamma(\eta,t,0)}{-\thb_0 +
 i\nu t^{-2}\Gamma(\eta,t,0)},
$$
which equals $\pi t^{-2} F(\eta, t,0) +O(\nu)$. So  $\lan I'_\nu, \Svm\ran=
\pi \int_{\Sigma_*}\frac{F(z)}{|z|}d_{\Sigma_*}z +O(\nu)\,.
$
The integral over the complement to $\Svm$ is 
$$
  \int_{\R^{2d}\setminus \Svm} \frac{F(x,y)\,dxdy}{x\cdot y +i \Gamma(x,y)}
 = \int_{\R^{2d}\setminus \Svm} \frac{F(x,y)\,dxdy}{x\cdot y } +o(1)
$$
as $\nu\to0$ (the integral in the r.h.s. is regular). 
In difference with  \eqref{I_s}  the last  integral is of the same order as the 
integral over $\Svm$.  So we  have that 
$$
I'_\nu=
\pi \int_{\Sigma_*}\frac{F(z)}{|z|}d_{\Sigma_*}z
 +
  \int_{\R^{2d}\setminus \Sv} \frac{F(x,y)\,dxdy}{x\cdot y } +o(1)\,,
$$
in agreement with  the estimate \eqref{Aa3}, applied to \eqref{another_integral}. 

\subsection{Integrals, coming from the three-waves interaction}
The three-waves interacting systems lead to integrals, similar to \eqref{heur}, where $\R^{3d}$ is
replaced by $\R^{2d}$ and the $\delta$-factor is replaced by $\delta^k_{k_1 k_2} \delta(\omega^k_{k_1 k_2})$,
which gives rise  to the algebraic set 
$$
 \{ (k_1, k_2): k_1+k_2 = k,\;\; |k_1|^2+|k_2|^2 = |k|^2\}
$$
(see \cite{Naz11}, Section 6).  I.e., $k_2 = k-k_1$, where $k_1 \in \{r\in\R^d: |r-\tfrac12 k|^2 = \tfrac14|k|^2\}$. 
Accordingly, in 
the variable $z=r-\tfrac12 k$ some constructions from the study of the three-waves interaction lead to
the integrals
$$
I'_\nu = \int_{\R^d} dz\, \frac{F(z)}{\omega(z)^2 + (\nu\G(z))^2}\,,\qquad d\ge2,
$$
with $\omega(z) = |z|^2 - \tfrac14|k|^2$. Now the quadric $\Sigma = \{\omega=0\}$ is a sphere, i.e. a
 smooth compact manifold. Denoting by $\eta$ a local coordinate on $\Sigma$ with a coordinate mapping
 $
 \eta \mapsto z(\eta) \in \Sigma
 $
 and   the volume form 
 $m(d\eta)$ we see that, similar to Section~\ref{s_3}, the local coordinate in the  vicinity $\Sv$ of $\Sigma$
  is $(\eta,\theta)$, $|\theta|<\theta_0$,    with the coordinate mapping
  $\,
  (\eta, \theta) \mapsto z(\eta) (1+\theta)\, 
  $
  and the 
 volume form $ (\tfrac12|k|)^{-1} \mu(\eta,\theta)m(d\eta)d\theta$, $\mu(\eta,0)\equiv0$. 
 The proof in Sections~\ref{s_71}--\ref{s4}
  simplifies  and leads to the asymptotic 
 $$
  \big| I'_\nu  - \pi \nu^{-1} \int_\Sigma \big( F(z)/\G(z)\big)dz \big| \le \text{Const},
 $$
 valid for $C^2$-functions $F$ and $\G$, satisfying some mild restriction.

\section{Appendix}
   
   Let $\phi(x)$ and $g(x)$ be smooth functions on $\R^n$ and $\phi$ has a  compact support.
   Consider the integral
   $$
   I(\lambda)= \int_{\R^n} \phi(x) e^{i\lambda g(x)}\,dx,\quad \lambda\ge1. 
   $$
   Assume that $g(x)$ has a unique  critical point  $x_0$, which is non-degenerate. 
   Then, by the stationary phase method, 
   \be\label{stph_thm}
     I(\lambda) = 
      \big(\frac{2\pi}{\la}\big)^{n/2}
      |\det g_{xx} (x_0)|^{-1/2} \phi(x_0) e^{i\lambda g(x_0) +(i\pi/4)\, \text{sgn}\, g_{xx}(x_0)  } 
     +R  \la^{-n/2 -1}\;\; 
   \ee
   for $\la\ge1$, 
   where  $R$ depends on $\|\phi\|_{C^2}$,   $\| g \|_{C^3}$, the measure of the support of $\phi$ 
    and on
   $\ 
   \sup_{x\in\text{supp} \,\phi} (|x-x_0| / |\nabla g(x)|) =: C^\#(g).
   $
   See  Section~7.7 of \cite{Hor}  and  Section~5 of  \cite{MF}. 
   
   If the functions $\phi$ and $g$ are not $C^\infty$--smooth, but $\phi\in C^2_0(\R^n)$ and $g\in C^3(\R^n)$, 
   then, approximating $\phi$ and $g$ by smooth functions and applying the result above we get from 
   \eqref{stph_thm}  that
   \be\label{stph_est}
      |I(\lambda) | \le C' \la^{-n/2}\qquad \forall\,\la\ge1\,,
   \ee
   with $C'$ depending  on  $\|\phi\|_{C^2}$,   $\| g \|_{C^3}$ and $C^\#(g)$. 
   \bigskip

   Now let  $f(x)\in C^2_0(\R^d)$ and    $g(x) \in C^3(\R^d)$ be such that 
$$
 |f|\le C,\;\; \meas(\text{\,supp} f ) \le C.
$$
Let an $x_0$ be the unique  critical point of $g(x)$ and 
\be\label{hess}
 C^{-1} \le |\det\text{Hess}\,g(x_0)|  \le C.
\ee

Consider the integral 
$$
\cI(\nu) = \int_{\R^d} \frac{f(x)}{\Gamma+i\nu^{-1} g(x)}\,dx =
\nu  \int_{\R^d} \frac{f(x)}{\nu\Gamma+i g(x)}\,dx
\,, \;\; 0<\nu\le1\,,
$$
where $ \Gamma$ is a positive constant. Let us write it as 
$$
\cI(\nu) = \int_{\R^d} \int_{-\infty}^0 f(x) e^{t(\Gamma +i\nu^{-1}  g(x))}\,dt\,dx =:  \int_{\R^d} \int_{-\infty}^0 F_\nu(t,x) \,dt\,dx =
\cI_1+\cI_2\,,
$$
where 
$$
 \cI_1=  \int_{\R^d} \int_{-\nu}^0 F_\nu(t,x) \,dt\,dx\,,\;\; \;\;
  \cI_2=
 \int_{\R^d} \int_{-\infty}^{-\nu} F_\nu(t,x) \,dt\,dx\,.
$$
Clearly, $|\cI_1|\le C^2\nu$.  To estimate $\cI_2$ consider the internal  integral 
$$
{J}(t) = e^{t\Gamma} 
 \int_{\R^d}  f(x) e^{-i\nu ^{-1} |t|g(x)}\,dx\,,
$$
and apply to it the stationary phase method with $\la=\nu^{-1}|t|\ge1$.
 By \eqref{stph_est} and \eqref{hess}, 
$|{J}(t)|$ is bounded by   $e^{t\Gamma }K_1(f,g)(\nu^{-1} |t|)^{-d/2}$. So
$$
|\cI_2 | = \int_{-\infty}^\nu J(t)\,dt
\le K_1(f,g)  \nu^{d/2} \int_{-\infty}^\nu | t|^{-d/2} e^{t\Gamma}\,dt 
\le K_2(f,g) \nu^{d/2} \nu^{-d/2 +1}\chi_d(\nu)
$$
since $\Gamma>0$, where $\chi_d(\nu)$ is defined in 
 \eqref{chi_d}. 
 
 Thus 
\be\label{Aa3}
|\cI(\nu)| \le |\cI_1| + |\cI_2 |\le 
 K(f,g)\, \nu\,\chi_d(\nu)\,.
\ee
The constant $K(f,g)$ depends on  $C$, $\| f \|_{C^2}$,   $\| g \|_{C^3}$ and $C^\#(g)$.


\begin{thebibliography}{99}
 \bibitem{MF}
 M. Dimassi, J. Sj\"ostrand, \textit{ Spectral Asymptotics in the Semi-Classical Limit}, 
 CUP 1999.


 \bibitem{Hor}
 L. H\"ormander 
  \textit{ The Analysis of Linear Partial Differential Equations. Vol.~1}, 
 Springer 1983. 

\bibitem{KM15}
S. Kuksin, A. Maiocchi,
\textit{Derivation of a wave kinetic equation from the
  resonant-averaged stochastic {NLS} equation}, Physica~D  \textbf{309}
   (2015), 65-70. 

\bibitem{Oberv}
S. Kuksin,
\textit{The Zakharov-Lvov stochastic model for the wave turbulence}, in   ``Interactions in Geophysical 
Fluids", Oberwolfach Report 39/2016 (2016), 37-39.


 
   
\bibitem{Naz11}
S. Nazarenko, \textit{Wave {T}urbulence}, Springer 2011.




    
\end{thebibliography}
\end{document}